\documentclass[a4paper, amsfonts, amssymb, amsmath, reprint, showkeys, nofootinbib, twoside, superscriptaddress, longbibliography]{revtex4-2}

\usepackage{amsmath}
\usepackage{amsfonts}
\usepackage{amssymb}
\usepackage{changes}
\usepackage[english]{babel}
\usepackage[utf8]{inputenc}
\usepackage{parskip}
\usepackage{changes}
\usepackage[pdftex, pdftitle={Article}, pdfauthor={Author}]{hyperref} 

\newcommand{\Icp}{\ensuremath{I_\mathrm{c}^+}}
\newcommand{\Icm}{\ensuremath{I_\mathrm{c}^-}}
\newcommand{\Ir}{\ensuremath{I_\mathrm{r}}}
\newcommand{\Iac}{\ensuremath{I_\mathrm{ac}}}
\newcommand{\Vout}{\ensuremath{V_\mathrm{out}}}
\newcommand{\Rn}{\ensuremath{R_\mathrm{n}}}
\newcommand{\MEuS}{\ensuremath{M_\mathrm{EuS}}}

\newcommand{\Iin}{\ensuremath{I_\mathrm{in}}}
\newcommand{\Idis}{\ensuremath{I_\mathrm{dis}}}

\begin{document}
\title{Highly Efficient Superconducting Diodes and Rectifiers for Quantum Circuitry}

\author{Josep Ingla-Ayn\'es}

    \altaffiliation{These authors contributed equally to this work.}
    \email{jingla@mit.edu}
    \affiliation{Francis Bitter Magnet Laboratory and Plasma Science and Fusion Center,
Massachusetts Institute of Technology, Cambridge, Massachusetts 02139, USA}

\author{Yasen Hou$^*$}

    \affiliation{Francis Bitter Magnet Laboratory and Plasma Science and Fusion Center,
Massachusetts Institute of Technology, Cambridge, Massachusetts 02139, USA}
\author{Sarah Wang}
    \affiliation{Francis Bitter Magnet Laboratory and Plasma Science and Fusion Center,
Massachusetts Institute of Technology, Cambridge, Massachusetts 02139, USA}
\author{En-De Chu}
    \affiliation{Department of Physics and Astronomy, University of California, Riverside, California 92521, USA}
    \author{Oleg A. Mukhanov}
    \affiliation{SEEQC, Inc., Elmsford, New York 10523, USA}
\author{Peng Wei}
    \affiliation{Department of Physics and Astronomy, University of California, Riverside, California 92521, USA}
\author{Jagadeesh S. Moodera}
\email{moodera@mit.edu}
    \affiliation{Francis Bitter Magnet Laboratory and Plasma Science and Fusion Center,
Massachusetts Institute of Technology, Cambridge, Massachusetts 02139, USA}

\affiliation{Department of Physics, Massachusetts Institute of Technology, Cambridge, Massachusetts 02139, USA}
\date{\today} 

\begin{abstract}
Superconducting electronics 
is essential for energy-efficient quantum and classical high-end computing applications. Towards this goal, non-reciprocal superconducting circuit elements, such as superconducting diodes (SDs) can fulfill many critical needs. SDs have been the subject of multiple studies, but integrating several SDs in a superconducting circuit remains a challenge. Here we implement the first SD bridge with multiple SDs exhibiting reproducible characteristics operating at temperatures of a few Kelvin. We demonstrate its functionality as a full wave rectifier using elemental superconductors and insulating ferromagnets, with efficiency up to 43\%, and ac to dc signal conversion capabilities at frequencies up to 40 kHz. Our results show a pathway with a highly scalable thin film platform for nonreciprocal superconducting circuits. They could significantly reduce energy consumption as well as decohering thermal and electromagnetic noise in quantum computing.
\end{abstract}


\maketitle
Over the last decades, computation operations have concentrated in data centers 
{that process large quantities of information} worldwide. According to the International Energy Agency \cite{IEA2024}, their total power consumption reached 460~TWh in 2022 and will increase to at least 620~TWh by 2026. In this context, the requirement for energy-efficient high-end computing is highly sought. 

Aided by the concentration of computing power {in data centers} and improvements in cryogenic cooling efficiency \cite{radebaugh2009}, superconducting electronics come to the forefront as a promising alternative for classical and quantum computing. Furthermore, its applications extend to astronomy detectors, magnetometry, and voltage standards \cite{braginski2019} and ongoing research expands them to dark matter detection \cite{dixit2021}. However, its full exploitation for high-end computing requires the optimization of different SC analogs of semiconducting devices. 
One of the critical long-standing requirements has been the need for the efficient delivery of dc bias current for superconducting energy-efficient rapid single flux quantum (ERSFQ) circuits \cite{kirichenko2011,mukhanov2011} which avoids the rise of total dc bias current with the number of cells. This problem limits the ERSFQ scalability to larger circuit complexities. This is also important for the millikelvin ERSFQ circuits designed to implement the qubit control and readout while minimizing thermal load and electromagnetic noise \cite{mukhanov2019, mcdermott2018}. {In this context, the recently rediscovered superconducting diodes (SDs) are superconductors (SCs) displaying nonreciprocal flow of charges and represent an ideal building block to be exploited for superconducting electronics \cite{braginski2019, cai2023, nadeem2023}, and are capable of solving the existing problems listed above.}

\begin{figure*}
	\centering
		\includegraphics[width=\textwidth]{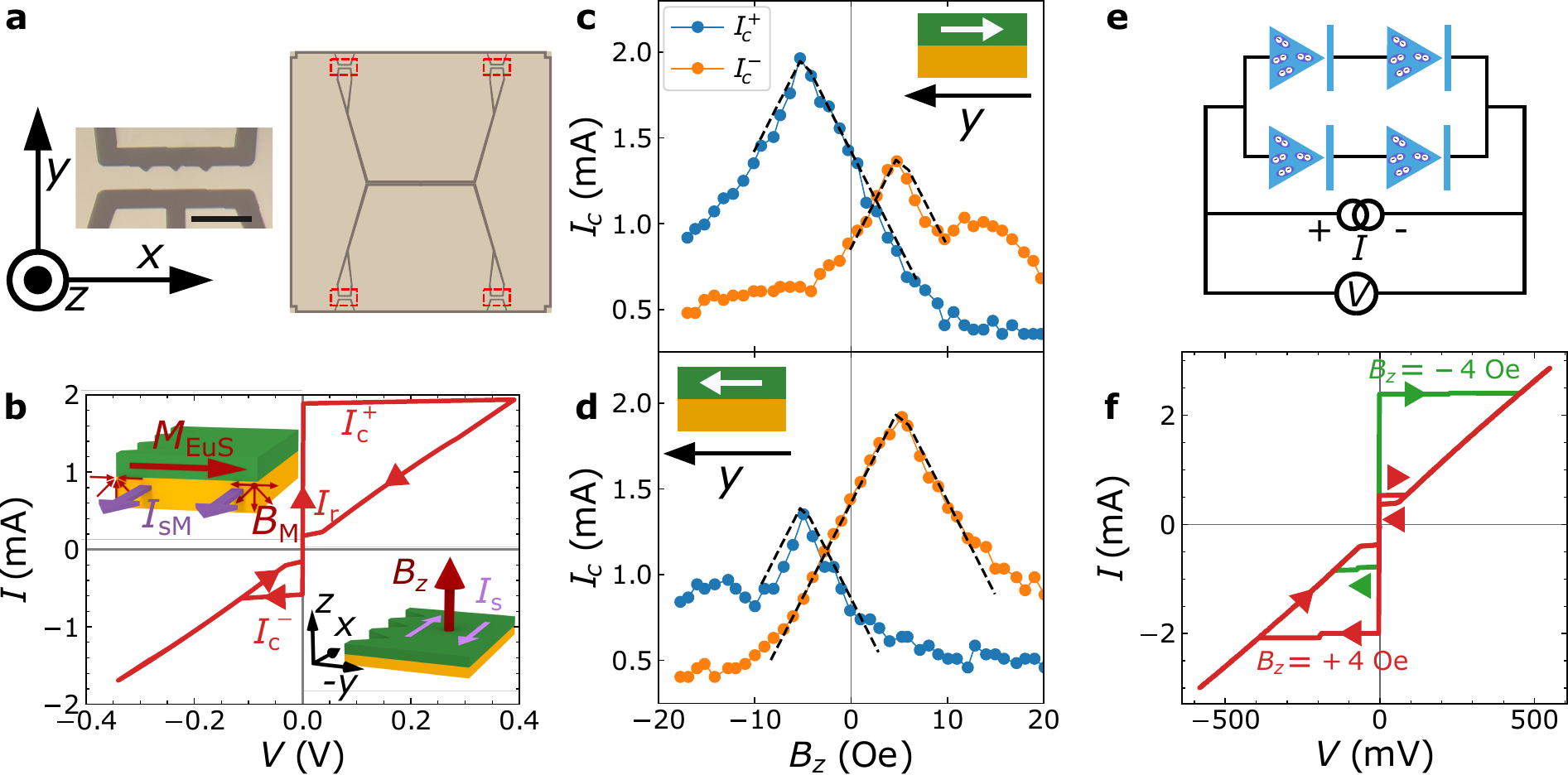}
	\caption{\textbf{Efficient and reproducible superconducting diode (SD) effect in V/EuS devices.} (a) The left panel is an optical microscope image of an SD where the dark areas have been etched. The scale bar is 20~$\mu$m. The right panel is a schematic of the superconducting rectifier with the same colors as the left image. Red rectangles highlight the SDs, and the outer square size is $1\times1$~mm$^2$. (b) $I$-$V$ characteristics of a single SD for $B_z=-5$~Oe and \MEuS{} along $-y$. The arrows indicate the sweep directions and the forward, reverse, and retrapping currents (\Icp{}, \Icm{}, and \Ir{}, respectively) are defined. The upper left inset illustrates the SD induced by stray magnetic fields from the EuS (green) acting on the V (orange). The lower right inset shows the SD effect induced by the out-of-plane magnetic field ($B_z$) in combination with edge asymmetry. (c) and (d) \Icp{} and \Icm{} of the SD in panel b as a function of $B_z$ after aligning the EuS magnetization as indicated by the insets. The black dashed lines are fits to the model described in the text. The remanence of the setup magnet is corrected assuming that the \Icp{} and \Icm{} peaks are centered around zero. The error bars corresponding to one standard deviation are smaller or equal to the dot size. (e) Schematic representation of a superconducting diode bridge with the circuit used to measure the SD effect. The SDs are sketched as blue diodes. (f) SD effect in a superconducting diode bridge at $B_z\approx\pm 4$~Oe. The measurement temperature is 1.7~K.}
	\label{Figure1}
\end{figure*}
\begin{figure*}
	\centering
		\includegraphics[width=0.9\textwidth]{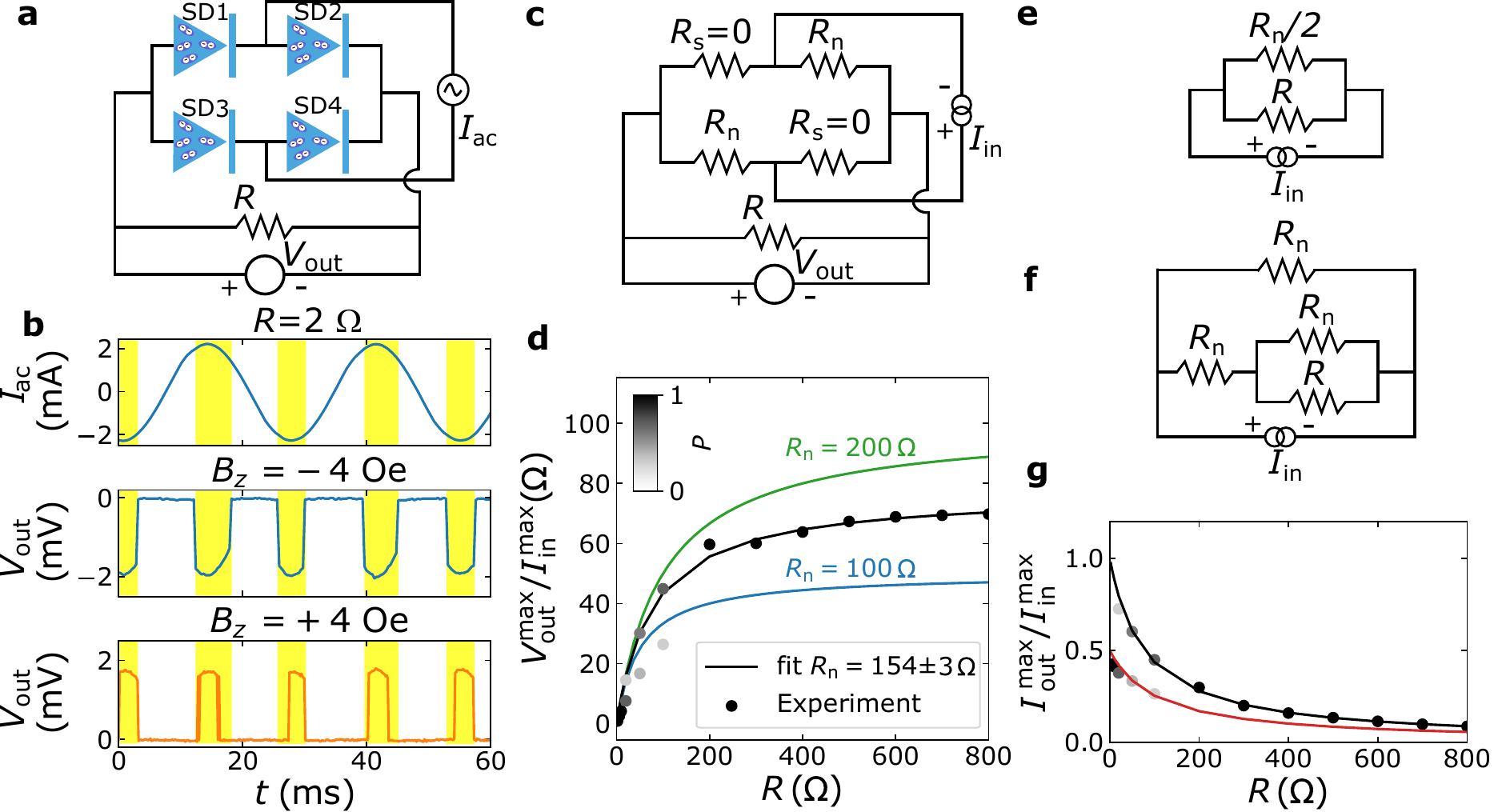}
	\caption{\textbf{Reversible operation and efficiency of a superconducting rectifier.} (a) Schematic of the measurement circuit used to measure the rectification of a superconducting diode bridge rectifier. $R$ is the resistance of the output load. (b) Experimental demonstration of reversible superconducting diode operation for $R=2\,\Omega$. The upper panel shows the input current (\Iac{}), measured for improved accuracy and smoothed using a moving average filter with a 7-point window. The middle and lower panels show the output voltages at $B_z=\pm4$~Oe, respectively. The yellow stripes highlight the time intervals where the reverse diodes are in the normal state ($|\Iac{}|\geq2\min(\Icp{},\,\Icm{})$). The \Vout{} amplitude is slightly lower in the lower panel due to a $\sim5$\% lower \Iac{} (not shown for simplicity), an offset of 0.1~mV has been corrected from \Vout{}, and the measurement temperature is 1.7~K. (c) Electrical circuit used to simulate the rectifier performance. The input current ($I_\mathrm{in}$) is constant and sufficiently large to turn SD2 and SD3 resistive  (see labels), acquiring a resistance \Rn{} while SD1 and SD4 remain superconducting. (d) Maximal output signal as a function of $R$. The dots are experimental values, and the lines are predictions of the resistor model in panel c for different \Rn{}. For $10<R<200\,\Omega$, \Vout{} peaks with two different amplitudes are observed. Their measurement probability ($P$) determines the color of the points according to the color bar. (e) Equivalent resistor circuit to panel c. (f) Same as panel e but considering that SD1 or SD4 become resistive. (g) Rectifier efficiency vs.~$R$ extracted from panel c. The black and red lines are obtained from the models in panels e and f, respectively with $\Rn{}=154\,\Omega$. The dots in panels d and g are larger than the error bars corresponding to one standard deviation and the uncertainty of \Rn{} in panel d corresponds to the fitting error.}
	\label{Figure2}
\end{figure*}
SDs are systems where inversion and time reversal symmetries are broken \cite{tokura2018}, a condition that can be achieved using different approaches. External magnetic fields \cite{villegas2003, ando2020, suri2022, hou2023,chahid2023}, ferromagnets \cite{narita2022,jiang2022, trahms2023,hou2023}, and trapping Abrikosov vortices in Josephson junctions \cite{golod2022} can break time-reversal symmetry. Moreover, inversion symmetry can be broken using materials with low crystallographic symmetry \cite{wakatsuki2017, narita2022, lin2022}, introducing device-scale asymmetries \cite{swartz1967,villegas2003, 1vodolazov2005,2vodolazov2005,moll2023,suri2022,hou2023}, or even applying inhomogeneous magnetic fields \cite{edwards1962, swartz1967}. Thanks to this versatility, a wide variety of approaches have been employed to realize SDs \cite{villegas2003, ando2020, suri2022, hou2023,wakatsuki2017, jiang2022, trahms2023, golod2022, wakatsuki2017, narita2022, lin2022, swartz1967, 1vodolazov2005,2vodolazov2005,moll2023, edwards1962, yasuda2019, itahashi2020, wu2022, jeon2022, baumgartner2022, ilic2022, kokkeler2022, yuan2022, davydova2022, pal2022,hope2021,chahid2023,zhao2023,ghosh2024,bozkurt2023,cayao2024,yerin2024,pal2018}, opening the door for potential applications.

 To enable the great synergy of SDs with existing superconducting computation technologies 
 and 
 {realize} the on-chip conversion of ac signals into dc \cite{braginski2019, mukhanov2011}, nonreciprocal superconductivity applications require practical circuits involving multiple SDs. For this purpose, superconducting thin films with engineered edge asymmetry represent an economically viable approach, as the growth and patterning of superconducting thin films can be easily scaled to realize complex SD-based circuits. 

Here we realize the first SD-based circuit: an SD bridge. We optimize the SD efficiency of the individual diodes up to $\pm$50\% by combining the effect of edge asymmetry in the SC vanadium (V) and stray fields from a ferromagnetic insulator (EuS) in thin-film bilayers of V/EuS. By patterning four practically identical SDs on the same superconducting film, we demonstrate SD bridges where the operational polarity can be reversed using magnetic fields. To showcase its implications for superconducting quantum technologies, we demonstrate its ability to convert ac signals up to 40 kHz into dc signals.

\begin{figure}[tb]
	\centering
		\includegraphics[width=0.35\textwidth]{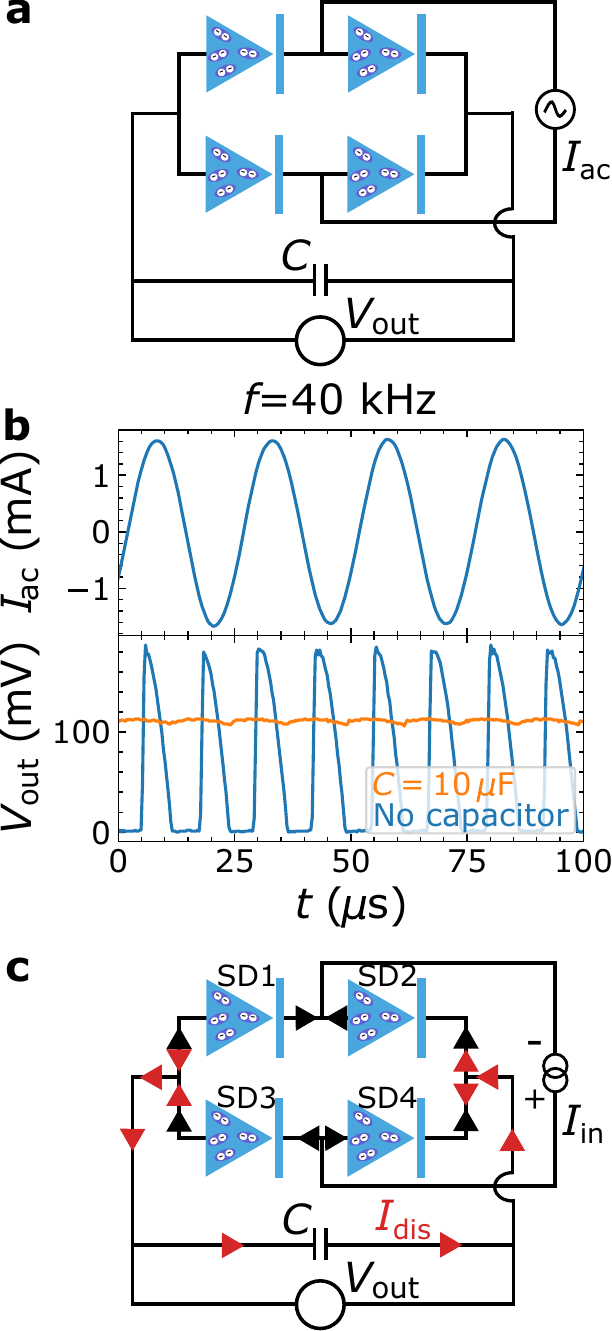}
	\caption{\textbf{Converting ac currents into dc signals using a superconducting rectifier.} (a) Schematics of the measurement circuit used to convert ac currents where a smoothing capacitor ($C$) has been introduced. (b) Superconducting rectifier operation at $f=40$~kHz and $C=10\,\mu$F effect. The measurement temperature is 1.7~K and \Iac{} has been smoothed using a zero-order Savitzky-Golay filter. (c) Schematic illustrating the effect of the discharging current (\Idis{}, red triangles) in combination with \Iin{} (black triangles) to enhance the rectifier efficiency (see text).}
	\label{Figure3}
\end{figure}
Our SDs are created growing 6-nm-thick V and 5-nm-thick EuS on sapphire in a molecular beam epitaxy system at a base pressure below $4\times10^{-10}$~Torr. The SDs are defined by e-beam lithography and ion milling. To maximize the SD efficiency, we designed FM/SC structures with asymmetrical edges \cite{1vodolazov2005, 2vodolazov2005, hou2023,suri2022}, as shown in Fig.~\ref{Figure1}a.  

Fig.~\ref{Figure1}b shows a 2-probe $I$-$V$ scan used to identify the forward and reverse critical currents (\Icp{} and \Icm{}, respectively) for increasing $|I|$ and the retrapping current (\Ir{}) for decreasing $|I|$. The difference between \Icp{} and \Icm{} determines the efficiency of the SD operation. In Fig.~\ref{Figure1}b, $\Icp{}\approx1.96$~mA, and $\Icm{}\approx-0.64$~mA. Thus, the efficiency $\eta=(\Icp{}-\Icm{})/(\Icp{}+\Icm{})\approx 51\%$, among the largest values reported in the literature. This result is obtained by aligning the EuS magnetization (\MEuS{}) along $-y$, creating the stray magnetic field $B_\mathrm{M}$ shown in Fig.~\ref{Figure1}b, left inset, which enhances \Icp{} and reduces \Icm{}. Following Ref.~\cite{hou2023}, we call this mechanism type C SD effect. In addition, we applied an out-of-plane magnetic field ($B_z=-5$~Oe) to enhance the SD rectification. As shown in the right inset of Fig.~\ref{Figure1}b, this induces an opposite Oersted current in both edges which, together with the edge asymmetry, results in a large SD effect\cite{2vodolazov2005, suri2022, hou2023}, which we call type A SD effect. The $B_z$-dependence of \Icp{} and \Icm{} is shown in Fig.~\ref{Figure2}c. The double V dependence of \Icp{} and \Icm{} is a consequence of the type A SD effect, and the higher \Icp{} is caused by the type C SD effect. This combination results in $\eta\approx51$\% for $B_z=-5$~Oe (see also Fig.~\ref{Figure2}b). Aligning the EuS magnetization towards $+y$ reverses the type C SD effect, enhancing \Icm{} and reducing \Icp{}, as shown in Fig.~\ref{Figure2}d. This effect results in $\eta\approx -50$\%, allowing the full reversal of the SD efficiency.

The enhancement of the diode efficiency by combining type A and type C diode effects is robust and reversible. Nearly identical dependence of \Icp{} and \Icm{} on \MEuS{} and $B_z$ was observed in 4 single diodes (Fig.~S1). Such a consistency enabled us to fabricate the SC rectifier circuit shown in the right panel of Fig.~\ref{Figure1}a, which consists of four single diodes set in the same direction (schematically shown in Fig.~\ref{Figure1}e). Typical two-probe $I$-$V$ scans of the rectifier with optimized efficiency are shown in Fig.~\ref{Figure1}f. With \MEuS{} aligned along $-y$, $\eta$ reaches a peak value of 50\% at $B_z = -4$~Oe (green curve). When \MEuS{} is reversed, the maximal $\eta\approx-47$\% is observed at $B_z = +4$~Oe (red curve). Very similar performance has been achieved in a second V/EuS superconducting rectifier (see Fig.~S7).

Under those optimized conditions, we exploit the functionality of the SD bridge using the full wave rectifier circuit shown in Fig.~\ref{Figure2}a \cite{horowitz1989}. 
The results from these measurements are shown in Fig.~\ref{Figure2}b for $R=2\,\Omega$, where $R$ is the load resistor. The upper panel shows the input ac current, which is kept smaller than $\max(\Icp{}, \Icm{})$ to prevent the forward SD to transition to the resistive state, and the middle and lower panels show the output voltage (\Vout{}) for $B_z=-4$ and $+4$~Oe, respectively. These figures demonstrate the operation of a reversible SD bridge with $|\Vout{}|>0$ when $|\Iac{}|>2\min(\Icp{}, \Icm{})$ (orange areas), i.e. the reverse SDs [SD2 and SD3 (SD1 and SD4) for positive (negative) \Iac{}, see Fig.~\ref{Figure2}a for the diode labels] are in the resistive state. The small discrepancy between the magnitude of \Vout{} in the middle and lower panels is due to a small difference in the \Iac{} amplitude.

The resistor model in Fig.~\ref{Figure2}c is used to understand the measured signals. For positive $\Iac{}=\Iin{}$, 
SD1 and SD4 are reversed. This means that for $\Iin{}>2\Icm{}$, they acquire a resistance \Rn{}. As shown in Fig.~\ref{Figure2}c, since SD2 and SD3 remain superconducting, the voltage drop across them is zero, resulting in a positive output voltage ($\Vout{}$). Reversing \Iac{} turns SD2 and SD3 resistive. Since SD1 and SD4 remain superconducting, there is no voltage drop across them and $\Vout{}$ remains positive. Note that, if $|\Iac{}|<2\min(\Icp{},\,\Icm{})$, the four SDs are SC and there is no current across $R$. This model assumes that $\Rn{}$ is the same for the four SDs.

We evaluate the efficiency of the rectifying operation in the \Iac{} range of Fig.~\ref{Figure2}c (SD2 and SD3 are resistive) by measuring \Vout{} for different values of $R$. Its maximal value is plotted in Fig.~\ref{Figure2}c. After increasing significantly with $R$ for $R<200\,\Omega$, the output signal saturates, indicating a parallel current flow. To understand this result, in Fig.~\ref{Figure2}e, we show the equivalent resistor circuit of Fig.~\ref{Figure2}c, which shows that the output resistor is now parallel with two resistive SDs. As a consequence, the output signal can be expressed as a function of $R$ using 
\begin{equation}
  \Vout{}/\Iin{}=\frac{R}{2R/R_n+1},
  \label{Equation1}
\end{equation}
 which is shown by the solid lines in Fig.~\ref{Figure2}d. Using Eq.~\ref{Equation1} to fit the measured data by least squares gives $R_n=156\,\Omega$, which is comparable with the $R_n\approx200\,\Omega$ obtained from the slope of the $I$-$V$ curves in Fig.~\ref{Figure1}f. We attribute this slight discrepancy to small differences in $R_n$ between the SDs. The current efficiency is represented in Fig.~\ref{Figure2}d, where the black line is the expected efficiency 
 \begin{equation}
    I_\mathrm{out}/I_\mathrm{in}=P_\mathrm{out}/P_\mathrm{in}=\frac{1}{2R/\Rn{}+1}, 
    \label{Equation2}
 \end{equation}
  extracted from Fig.~\ref{Figure2}e with $R_n=156\,\Omega$, where $I_\mathrm{out(in)}$ and $P_\mathrm{out(in)}$ are the output (input) current and power, respectively. The black dots are obtained from the experimental data and agree well with the model for high $R$. However, for $R<200\,\Omega$, the data in Fig.~\ref{Figure2}e shows two sets of peaks with a factor of two difference in the amplitude (Fig.~S3) which do not agree with Eq.~\ref{Equation2}. The dot color indicates the probability ($P$) by which they occur in the measurements according to the color bar in Fig.~\ref{Figure2}d.
  We explain these lower-amplitude peaks considering that either SD1 or SD4 becomes normal. This results in the equivalent circuit shown in Fig.~\ref{Figure2}f, which gives the following expression for the current efficiency
  \begin{equation}
    I_\mathrm{out}/I_\mathrm{in}=\frac{1}{2\left(\frac{3R}{2\Rn{}}+1\right)}. 
    \label{Equation3}
 \end{equation}
 When $R\ll \Rn{}$, the efficiency given by Eq.~\ref{Equation3} is two times smaller than Eq.~\ref{Equation2}. This is caused by the increase in the resistance of the current path that goes through the load resistor $R$, which penalizes the circuit efficiency even for $R=0$. Using $R_n=156\,\Omega$, Eq.~\ref{Equation3} results in the red line in Fig.~\ref{Figure2}d, which shows good agreement with the experiment at low $R$. This result is independent of which SD is turning normal (SD1 or SD4). 
Two possible reasons can explain this premature transition to the resistive state: 1. Heating induced by SD2 and SD3 decreases \Icp{} of SD1 and SD4. 2. Small differences in \Icp{} between the SDs in the bridge. These effects do not occur for higher $R$ because the current load on the forward diodes decreases with increasing $R$. Nevertheless, the maximal efficiency achieved is 42$\pm$2\%, a remarkable metric for such a new device.

To fully exploit our superconducting rectifier to convert ac into dc signals at cryogenic temperatures, we couple it to a smoothing capacitor $C$ (Fig.~\ref{Figure3}a) \cite{horowitz1989}. When $C=10\,\mu$F and we apply \Iac{} at a frequency of 40~kHz, a value much larger than $2/(C R_n)$, the oscillations of \Vout{} are attenuated and a flat signal arises (Fig.~\ref{Figure3}b, orange curve). Thus, we demonstrate the successful conversion of the supplied ac signal into a dc voltage by our SD rectifier circuit. 
The conversion of ac into dc signals has been observed for frequencies up to 40~kHz. Above this frequency the amplitude of the dc signal drops. Without the capacitor, the rectifier circuit is functional up to 100~kHz (Fig.~S4). We attribute this limitation to the cryostat wiring, which provides an unwanted coupling between the wires. 
Analyzing the dc signal magnitude, we observe that it is approximately 59\% of the maximum peak signal. We calculate the expected dc signal magnitude by simulating the diode bridge as a time-dependent voltage source connected in parallel with a resistor $R_\mathrm{eq}=\Rn{}/2$ and a capacitor $C$ (see Fig.~{S6}). The result is approximately 45\% of the maximum \Vout{}. We attribute the discrepancy to a rectifier efficiency enhancement induced by the capacitor discharge current (\Idis{}). As illustrated in Fig.~\ref{Figure3}c, when the input current \Iin{} is positive and $\Iin{}\Rn{}/2<\Vout{}$, the total current through SD2 and SD3 is $\Iin{}+\Idis{}$ while for SD1 and SD4 it is $\Iin{}-\Idis{}$. This implies that SD2 and SD3 become resistive for lower \Iin{}. Note that the average \Vout{} in the most optimal case (\Ir{}$=\Icm{}=0$) is $2/\pi\approx63.7$\% of the peak value without $C$. This effect has been measured at different frequencies in Fig.~S5. Figure~\ref{Figure3}b also shows a phase shift of \Vout{} when introducing $C$. The $C$-smoothed \Vout{} (orange) is expected to increase when the $C=0$ \Vout{} (blue) is larger and to decrease otherwise.

The versatility of our devices is further demonstrated by showing the performance of an Nb-based superconducting rectifier (see Fig.~S8).

To conclude, by combining type A and type C SD effects in V/EuS thin films, we have demonstrated robust single diodes with enhanced efficiency and tunability. With the superb efficiency and reproducibility achieved, we constructed the first superconducting rectifiers operating with frequencies up to 40~kHz and efficiencies up to 42$\pm$2\%. The integration of SC rectifiers in SC electronic chips is a key step towards the scaling of ERSFQ-based quantum and classical computers \cite{braginski2019}, which would also benefit new sensing applications \cite{dixit2021}. In addition, DC power delivery is not the only problem that SD circuits can tackle. Circulators are ubiquitous devices in the majority of quantum computing systems. The existing devices are bulky, expensive 3D objects increasing quantum system complexity and cost. Their integration in SD promises to boost quantum computation \cite{chapman2017}.

\emph{Note added:} During the submission process of this work we became aware of a related work on superconducting rectifiers \cite{castellani2024}.

\section{Acknowledgements}
We acknowledge Dr. M. B. Ketchen, Dr. A. Kirichenko, and A. Gupta for insightful discussions and M. Mondol for technical assistance. This work was supported by Air Force Office of Sponsored Research (FA9550-23-1-0004 DEF), Office of Naval Research (N00014-20-1-2306), National Science Foundation (NSF-DMR 2218550 and 1231319); Army Research Office (W911NF-20-2-0061, DURIP W911NF-20-1-0074). EC and PW acknowledge the NSF CAREER DMR-2046648. This work was carried out in part through the use of MIT.nano's facilities.
\section{Data availability}
The data supporting the findings of this study are available from the corresponding authors upon request.
\section{Author contributions}
JIA, YH, and JSM conceived and designed the study with input from OM. YH grew the V/EuS films, EDC and PW grew the Nb/Au films, and JIA fabricated the devices. JIA performed the measurements. YH and SW assisted with the measurements. All authors contributed to the manuscript.
\bibliography{bibliography}

\begin{thebibliography}{47}%
\makeatletter
\providecommand \@ifxundefined [1]{%
 \@ifx{#1\undefined}
}%
\providecommand \@ifnum [1]{%
 \ifnum #1\expandafter \@firstoftwo
 \else \expandafter \@secondoftwo
 \fi
}%
\providecommand \@ifx [1]{%
 \ifx #1\expandafter \@firstoftwo
 \else \expandafter \@secondoftwo
 \fi
}%
\providecommand \natexlab [1]{#1}%
\providecommand \enquote  [1]{``#1''}%
\providecommand \bibnamefont  [1]{#1}%
\providecommand \bibfnamefont [1]{#1}%
\providecommand \citenamefont [1]{#1}%
\providecommand \href@noop [0]{\@secondoftwo}%
\providecommand \href [0]{\begingroup \@sanitize@url \@href}%
\providecommand \@href[1]{\@@startlink{#1}\@@href}%
\providecommand \@@href[1]{\endgroup#1\@@endlink}%
\providecommand \@sanitize@url [0]{\catcode `\\12\catcode `\$12\catcode `\&12\catcode `\#12\catcode `\^12\catcode `\_12\catcode `\%12\relax}%
\providecommand \@@startlink[1]{}%
\providecommand \@@endlink[0]{}%
\providecommand \url  [0]{\begingroup\@sanitize@url \@url }%
\providecommand \@url [1]{\endgroup\@href {#1}{\urlprefix }}%
\providecommand \urlprefix  [0]{URL }%
\providecommand \Eprint [0]{\href }%
\providecommand \doibase [0]{https://doi.org/}%
\providecommand \selectlanguage [0]{\@gobble}%
\providecommand \bibinfo  [0]{\@secondoftwo}%
\providecommand \bibfield  [0]{\@secondoftwo}%
\providecommand \translation [1]{[#1]}%
\providecommand \BibitemOpen [0]{}%
\providecommand \bibitemStop [0]{}%
\providecommand \bibitemNoStop [0]{.\EOS\space}%
\providecommand \EOS [0]{\spacefactor3000\relax}%
\providecommand \BibitemShut  [1]{\csname bibitem#1\endcsname}%
\let\auto@bib@innerbib\@empty
\bibitem [{\citenamefont {Çam}\ \emph {et~al.}(2024)\citenamefont {Çam}, \citenamefont {Hungerford}, \citenamefont {Schoch}, \citenamefont {Miranda},\ and\ \citenamefont {de~León.}}]{IEA2024}%
  \BibitemOpen
  \bibfield  {author} {\bibinfo {author} {\bibfnamefont {E.}~\bibnamefont {Çam}}, \bibinfo {author} {\bibfnamefont {Z.}~\bibnamefont {Hungerford}}, \bibinfo {author} {\bibfnamefont {N.}~\bibnamefont {Schoch}}, \bibinfo {author} {\bibfnamefont {F.~P.}\ \bibnamefont {Miranda}},\ and\ \bibinfo {author} {\bibfnamefont {C.~D.~Y.}\ \bibnamefont {de~León.}},\ }\href {https://www.iea.org/reports/electricity-2024} {\bibinfo {title} {Electricity 2024 analysis and forecast to 2026}},\ \bibinfo {howpublished} {INTERNATIONAL ENERGY AGENCY} (\bibinfo {year} {2024})\BibitemShut {NoStop}%
\bibitem [{\citenamefont {Radebaugh}(2009)}]{radebaugh2009}%
  \BibitemOpen
  \bibfield  {author} {\bibinfo {author} {\bibfnamefont {R.}~\bibnamefont {Radebaugh}},\ }\bibfield  {title} {\bibinfo {title} {Cryocoolers: the state of the art and recent developments},\ }\href {https://iopscience.iop.org/article/10.1088/0953-8984/21/16/164219} {\bibfield  {journal} {\bibinfo  {journal} {Journal of Physics: Condensed Matter}\ }\textbf {\bibinfo {volume} {21}},\ \bibinfo {pages} {164219} (\bibinfo {year} {2009})}\BibitemShut {NoStop}%
\bibitem [{\citenamefont {Braginski}(2019)}]{braginski2019}%
  \BibitemOpen
  \bibfield  {author} {\bibinfo {author} {\bibfnamefont {A.~I.}\ \bibnamefont {Braginski}},\ }\bibfield  {title} {\bibinfo {title} {Superconductor electronics: Status and outlook},\ }\href {https://link.springer.com/article/10.1007/s10948-018-4884-4} {\bibfield  {journal} {\bibinfo  {journal} {Journal of superconductivity and novel magnetism}\ }\textbf {\bibinfo {volume} {32}},\ \bibinfo {pages} {23} (\bibinfo {year} {2019})}\BibitemShut {NoStop}%
\bibitem [{\citenamefont {Dixit}\ \emph {et~al.}(2021)\citenamefont {Dixit}, \citenamefont {Chakram}, \citenamefont {He}, \citenamefont {Agrawal}, \citenamefont {Naik}, \citenamefont {Schuster},\ and\ \citenamefont {Chou}}]{dixit2021}%
  \BibitemOpen
  \bibfield  {author} {\bibinfo {author} {\bibfnamefont {A.~V.}\ \bibnamefont {Dixit}}, \bibinfo {author} {\bibfnamefont {S.}~\bibnamefont {Chakram}}, \bibinfo {author} {\bibfnamefont {K.}~\bibnamefont {He}}, \bibinfo {author} {\bibfnamefont {A.}~\bibnamefont {Agrawal}}, \bibinfo {author} {\bibfnamefont {R.~K.}\ \bibnamefont {Naik}}, \bibinfo {author} {\bibfnamefont {D.~I.}\ \bibnamefont {Schuster}},\ and\ \bibinfo {author} {\bibfnamefont {A.}~\bibnamefont {Chou}},\ }\bibfield  {title} {\bibinfo {title} {Searching for dark matter with a superconducting qubit},\ }\href {https://journals.aps.org/prl/abstract/10.1103/PhysRevLett.126.141302} {\bibfield  {journal} {\bibinfo  {journal} {Physical Review Letters}\ }\textbf {\bibinfo {volume} {126}},\ \bibinfo {pages} {141302} (\bibinfo {year} {2021})}\BibitemShut {NoStop}%
\bibitem [{\citenamefont {Kirichenko}\ \emph {et~al.}(2011)\citenamefont {Kirichenko}, \citenamefont {Sarwana},\ and\ \citenamefont {Kirichenko}}]{kirichenko2011}%
  \BibitemOpen
  \bibfield  {author} {\bibinfo {author} {\bibfnamefont {D.}~\bibnamefont {Kirichenko}}, \bibinfo {author} {\bibfnamefont {S.}~\bibnamefont {Sarwana}},\ and\ \bibinfo {author} {\bibfnamefont {A.}~\bibnamefont {Kirichenko}},\ }\bibfield  {title} {\bibinfo {title} {Zero static power dissipation biasing of rsfq circuits},\ }\href {https://ieeexplore.ieee.org/document/5688194} {\bibfield  {journal} {\bibinfo  {journal} {IEEE Transactions on Applied Superconductivity}\ }\textbf {\bibinfo {volume} {21}},\ \bibinfo {pages} {776} (\bibinfo {year} {2011})}\BibitemShut {NoStop}%
\bibitem [{\citenamefont {Mukhanov}(2011)}]{mukhanov2011}%
  \BibitemOpen
  \bibfield  {author} {\bibinfo {author} {\bibfnamefont {O.~A.}\ \bibnamefont {Mukhanov}},\ }\bibfield  {title} {\bibinfo {title} {Energy-efficient single flux quantum technology},\ }\href {https://ieeexplore.ieee.org/document/5682046} {\bibfield  {journal} {\bibinfo  {journal} {IEEE Transactions on Applied Superconductivity}\ }\textbf {\bibinfo {volume} {21}},\ \bibinfo {pages} {760} (\bibinfo {year} {2011})}\BibitemShut {NoStop}%
\bibitem [{\citenamefont {Mukhanov}\ \emph {et~al.}(2019)\citenamefont {Mukhanov}, \citenamefont {Kirichenko}, \citenamefont {Howington}, \citenamefont {Walter}, \citenamefont {Hutchings}, \citenamefont {Vernik}, \citenamefont {Yohannes}, \citenamefont {Dodge}, \citenamefont {Ballard}, \citenamefont {Plourde} \emph {et~al.}}]{mukhanov2019}%
  \BibitemOpen
  \bibfield  {author} {\bibinfo {author} {\bibfnamefont {O.}~\bibnamefont {Mukhanov}}, \bibinfo {author} {\bibfnamefont {A.}~\bibnamefont {Kirichenko}}, \bibinfo {author} {\bibfnamefont {C.}~\bibnamefont {Howington}}, \bibinfo {author} {\bibfnamefont {J.}~\bibnamefont {Walter}}, \bibinfo {author} {\bibfnamefont {M.}~\bibnamefont {Hutchings}}, \bibinfo {author} {\bibfnamefont {I.}~\bibnamefont {Vernik}}, \bibinfo {author} {\bibfnamefont {D.}~\bibnamefont {Yohannes}}, \bibinfo {author} {\bibfnamefont {K.}~\bibnamefont {Dodge}}, \bibinfo {author} {\bibfnamefont {A.}~\bibnamefont {Ballard}}, \bibinfo {author} {\bibfnamefont {B.}~\bibnamefont {Plourde}}, \emph {et~al.},\ }\bibfield  {title} {\bibinfo {title} {Scalable quantum computing infrastructure based on superconducting electronics},\ }in\ \href {https://ieeexplore.ieee.org/document/8993634} {\emph {\bibinfo {booktitle} {2019 IEEE International Electron Devices Meeting (IEDM)}}}\ (\bibinfo {organization} {IEEE},\ \bibinfo {year} {2019})\ pp.\ \bibinfo {pages}
  {31--2}\BibitemShut {NoStop}%
\bibitem [{\citenamefont {McDermott}\ \emph {et~al.}(2018)\citenamefont {McDermott}, \citenamefont {Vavilov}, \citenamefont {Plourde}, \citenamefont {Wilhelm}, \citenamefont {Liebermann}, \citenamefont {Mukhanov},\ and\ \citenamefont {Ohki}}]{mcdermott2018}%
  \BibitemOpen
  \bibfield  {author} {\bibinfo {author} {\bibfnamefont {R.}~\bibnamefont {McDermott}}, \bibinfo {author} {\bibfnamefont {M.}~\bibnamefont {Vavilov}}, \bibinfo {author} {\bibfnamefont {B.}~\bibnamefont {Plourde}}, \bibinfo {author} {\bibfnamefont {F.}~\bibnamefont {Wilhelm}}, \bibinfo {author} {\bibfnamefont {P.}~\bibnamefont {Liebermann}}, \bibinfo {author} {\bibfnamefont {O.}~\bibnamefont {Mukhanov}},\ and\ \bibinfo {author} {\bibfnamefont {T.}~\bibnamefont {Ohki}},\ }\bibfield  {title} {\bibinfo {title} {Quantum--classical interface based on single flux quantum digital logic},\ }\href {https://iopscience.iop.org/article/10.1088/2058-9565/aaa3a0} {\bibfield  {journal} {\bibinfo  {journal} {Quantum science and technology}\ }\textbf {\bibinfo {volume} {3}},\ \bibinfo {pages} {024004} (\bibinfo {year} {2018})}\BibitemShut {NoStop}%
\bibitem [{\citenamefont {Cai}\ \emph {et~al.}(2023)\citenamefont {Cai}, \citenamefont {{\v{Z}}uti{\'c}},\ and\ \citenamefont {Han}}]{cai2023}%
  \BibitemOpen
  \bibfield  {author} {\bibinfo {author} {\bibfnamefont {R.}~\bibnamefont {Cai}}, \bibinfo {author} {\bibfnamefont {I.}~\bibnamefont {{\v{Z}}uti{\'c}}},\ and\ \bibinfo {author} {\bibfnamefont {W.}~\bibnamefont {Han}},\ }\bibfield  {title} {\bibinfo {title} {Superconductor/ferromagnet heterostructures: a platform for superconducting spintronics and quantum computation},\ }\href {https://onlinelibrary.wiley.com/doi/abs/10.1002/qute.202200080} {\bibfield  {journal} {\bibinfo  {journal} {Advanced Quantum Technologies}\ }\textbf {\bibinfo {volume} {6}},\ \bibinfo {pages} {2200080} (\bibinfo {year} {2023})}\BibitemShut {NoStop}%
\bibitem [{\citenamefont {Nadeem}\ \emph {et~al.}(2023)\citenamefont {Nadeem}, \citenamefont {Fuhrer},\ and\ \citenamefont {Wang}}]{nadeem2023}%
  \BibitemOpen
  \bibfield  {author} {\bibinfo {author} {\bibfnamefont {M.}~\bibnamefont {Nadeem}}, \bibinfo {author} {\bibfnamefont {M.~S.}\ \bibnamefont {Fuhrer}},\ and\ \bibinfo {author} {\bibfnamefont {X.}~\bibnamefont {Wang}},\ }\bibfield  {title} {\bibinfo {title} {The superconducting diode effect},\ }\href {https://www.nature.com/articles/s42254-023-00632-w} {\bibfield  {journal} {\bibinfo  {journal} {Nature Reviews Physics}\ }\textbf {\bibinfo {volume} {5}},\ \bibinfo {pages} {558} (\bibinfo {year} {2023})}\BibitemShut {NoStop}%
\bibitem [{\citenamefont {Tokura}\ and\ \citenamefont {Nagaosa}(2018)}]{tokura2018}%
  \BibitemOpen
  \bibfield  {author} {\bibinfo {author} {\bibfnamefont {Y.}~\bibnamefont {Tokura}}\ and\ \bibinfo {author} {\bibfnamefont {N.}~\bibnamefont {Nagaosa}},\ }\bibfield  {title} {\bibinfo {title} {Nonreciprocal responses from non-centrosymmetric quantum materials},\ }\href {https://www.nature.com/articles/s41467-018-05759-4} {\bibfield  {journal} {\bibinfo  {journal} {Nature Communications}\ }\textbf {\bibinfo {volume} {9}},\ \bibinfo {pages} {3740} (\bibinfo {year} {2018})}\BibitemShut {NoStop}%
\bibitem [{\citenamefont {Villegas}\ \emph {et~al.}(2003)\citenamefont {Villegas}, \citenamefont {Savel'ev}, \citenamefont {Nori}, \citenamefont {Gonzalez}, \citenamefont {Anguita}, \citenamefont {Garcia},\ and\ \citenamefont {Vicent}}]{villegas2003}%
  \BibitemOpen
  \bibfield  {author} {\bibinfo {author} {\bibfnamefont {J.}~\bibnamefont {Villegas}}, \bibinfo {author} {\bibfnamefont {S.}~\bibnamefont {Savel'ev}}, \bibinfo {author} {\bibfnamefont {F.}~\bibnamefont {Nori}}, \bibinfo {author} {\bibfnamefont {E.}~\bibnamefont {Gonzalez}}, \bibinfo {author} {\bibfnamefont {J.}~\bibnamefont {Anguita}}, \bibinfo {author} {\bibfnamefont {R.}~\bibnamefont {Garcia}},\ and\ \bibinfo {author} {\bibfnamefont {J.}~\bibnamefont {Vicent}},\ }\bibfield  {title} {\bibinfo {title} {A superconducting reversible rectifier that controls the motion of magnetic flux quanta},\ }\href {https://www.science.org/doi/10.1126/science.1090390} {\bibfield  {journal} {\bibinfo  {journal} {science}\ }\textbf {\bibinfo {volume} {302}},\ \bibinfo {pages} {1188} (\bibinfo {year} {2003})}\BibitemShut {NoStop}%
\bibitem [{\citenamefont {Ando}\ \emph {et~al.}(2020)\citenamefont {Ando}, \citenamefont {Miyasaka}, \citenamefont {Li}, \citenamefont {Ishizuka}, \citenamefont {Arakawa}, \citenamefont {Shiota}, \citenamefont {Moriyama}, \citenamefont {Yanase},\ and\ \citenamefont {Ono}}]{ando2020}%
  \BibitemOpen
  \bibfield  {author} {\bibinfo {author} {\bibfnamefont {F.}~\bibnamefont {Ando}}, \bibinfo {author} {\bibfnamefont {Y.}~\bibnamefont {Miyasaka}}, \bibinfo {author} {\bibfnamefont {T.}~\bibnamefont {Li}}, \bibinfo {author} {\bibfnamefont {J.}~\bibnamefont {Ishizuka}}, \bibinfo {author} {\bibfnamefont {T.}~\bibnamefont {Arakawa}}, \bibinfo {author} {\bibfnamefont {Y.}~\bibnamefont {Shiota}}, \bibinfo {author} {\bibfnamefont {T.}~\bibnamefont {Moriyama}}, \bibinfo {author} {\bibfnamefont {Y.}~\bibnamefont {Yanase}},\ and\ \bibinfo {author} {\bibfnamefont {T.}~\bibnamefont {Ono}},\ }\bibfield  {title} {\bibinfo {title} {Observation of superconducting diode effect},\ }\href {https://www.nature.com/articles/s41586-020-2590-4} {\bibfield  {journal} {\bibinfo  {journal} {Nature}\ }\textbf {\bibinfo {volume} {584}},\ \bibinfo {pages} {373} (\bibinfo {year} {2020})}\BibitemShut {NoStop}%
\bibitem [{\citenamefont {Suri}\ \emph {et~al.}(2022)\citenamefont {Suri}, \citenamefont {Kamra}, \citenamefont {Meier}, \citenamefont {Kronseder}, \citenamefont {Belzig}, \citenamefont {Back},\ and\ \citenamefont {Strunk}}]{suri2022}%
  \BibitemOpen
  \bibfield  {author} {\bibinfo {author} {\bibfnamefont {D.}~\bibnamefont {Suri}}, \bibinfo {author} {\bibfnamefont {A.}~\bibnamefont {Kamra}}, \bibinfo {author} {\bibfnamefont {T.~N.}\ \bibnamefont {Meier}}, \bibinfo {author} {\bibfnamefont {M.}~\bibnamefont {Kronseder}}, \bibinfo {author} {\bibfnamefont {W.}~\bibnamefont {Belzig}}, \bibinfo {author} {\bibfnamefont {C.~H.}\ \bibnamefont {Back}},\ and\ \bibinfo {author} {\bibfnamefont {C.}~\bibnamefont {Strunk}},\ }\bibfield  {title} {\bibinfo {title} {Non-reciprocity of vortex-limited critical current in conventional superconducting micro-bridges},\ }\href {https://pubs.aip.org/aip/apl/article/121/10/102601/2834233/Non-reciprocity-of-vortex-limited-critical-current} {\bibfield  {journal} {\bibinfo  {journal} {Applied Physics Letters}\ }\textbf {\bibinfo {volume} {121}} (\bibinfo {year} {2022})}\BibitemShut {NoStop}%
\bibitem [{\citenamefont {Hou}\ \emph {et~al.}(2023)\citenamefont {Hou}, \citenamefont {Nichele}, \citenamefont {Chi}, \citenamefont {Lodesani}, \citenamefont {Wu}, \citenamefont {Ritter}, \citenamefont {Haxell}, \citenamefont {Davydova}, \citenamefont {Ili{\'c}}, \citenamefont {Glezakou-Elbert} \emph {et~al.}}]{hou2023}%
  \BibitemOpen
  \bibfield  {author} {\bibinfo {author} {\bibfnamefont {Y.}~\bibnamefont {Hou}}, \bibinfo {author} {\bibfnamefont {F.}~\bibnamefont {Nichele}}, \bibinfo {author} {\bibfnamefont {H.}~\bibnamefont {Chi}}, \bibinfo {author} {\bibfnamefont {A.}~\bibnamefont {Lodesani}}, \bibinfo {author} {\bibfnamefont {Y.}~\bibnamefont {Wu}}, \bibinfo {author} {\bibfnamefont {M.~F.}\ \bibnamefont {Ritter}}, \bibinfo {author} {\bibfnamefont {D.~Z.}\ \bibnamefont {Haxell}}, \bibinfo {author} {\bibfnamefont {M.}~\bibnamefont {Davydova}}, \bibinfo {author} {\bibfnamefont {S.}~\bibnamefont {Ili{\'c}}}, \bibinfo {author} {\bibfnamefont {O.}~\bibnamefont {Glezakou-Elbert}}, \emph {et~al.},\ }\bibfield  {title} {\bibinfo {title} {Ubiquitous superconducting diode effect in superconductor thin films},\ }\href {https://journals.aps.org/prl/abstract/10.1103/PhysRevLett.131.027001} {\bibfield  {journal} {\bibinfo  {journal} {Physical Review Letters}\ }\textbf {\bibinfo {volume} {131}},\ \bibinfo {pages} {027001} (\bibinfo {year}
  {2023})}\BibitemShut {NoStop}%
\bibitem [{\citenamefont {Chahid}\ \emph {et~al.}(2023)\citenamefont {Chahid}, \citenamefont {Teknowijoyo}, \citenamefont {Mowgood},\ and\ \citenamefont {Gulian}}]{chahid2023}%
  \BibitemOpen
  \bibfield  {author} {\bibinfo {author} {\bibfnamefont {S.}~\bibnamefont {Chahid}}, \bibinfo {author} {\bibfnamefont {S.}~\bibnamefont {Teknowijoyo}}, \bibinfo {author} {\bibfnamefont {I.}~\bibnamefont {Mowgood}},\ and\ \bibinfo {author} {\bibfnamefont {A.}~\bibnamefont {Gulian}},\ }\bibfield  {title} {\bibinfo {title} {{High-frequency diode effect in superconducting Nb$_3$Sn microbridges}},\ }\href {https://journals.aps.org/prb/abstract/10.1103/PhysRevB.107.054506} {\bibfield  {journal} {\bibinfo  {journal} {Physical Review B}\ }\textbf {\bibinfo {volume} {107}},\ \bibinfo {pages} {054506} (\bibinfo {year} {2023})}\BibitemShut {NoStop}%
\bibitem [{\citenamefont {Narita}\ \emph {et~al.}(2022)\citenamefont {Narita}, \citenamefont {Ishizuka}, \citenamefont {Kawarazaki}, \citenamefont {Kan}, \citenamefont {Shiota}, \citenamefont {Moriyama}, \citenamefont {Shimakawa}, \citenamefont {Ognev}, \citenamefont {Samardak}, \citenamefont {Yanase} \emph {et~al.}}]{narita2022}%
  \BibitemOpen
  \bibfield  {author} {\bibinfo {author} {\bibfnamefont {H.}~\bibnamefont {Narita}}, \bibinfo {author} {\bibfnamefont {J.}~\bibnamefont {Ishizuka}}, \bibinfo {author} {\bibfnamefont {R.}~\bibnamefont {Kawarazaki}}, \bibinfo {author} {\bibfnamefont {D.}~\bibnamefont {Kan}}, \bibinfo {author} {\bibfnamefont {Y.}~\bibnamefont {Shiota}}, \bibinfo {author} {\bibfnamefont {T.}~\bibnamefont {Moriyama}}, \bibinfo {author} {\bibfnamefont {Y.}~\bibnamefont {Shimakawa}}, \bibinfo {author} {\bibfnamefont {A.~V.}\ \bibnamefont {Ognev}}, \bibinfo {author} {\bibfnamefont {A.~S.}\ \bibnamefont {Samardak}}, \bibinfo {author} {\bibfnamefont {Y.}~\bibnamefont {Yanase}}, \emph {et~al.},\ }\bibfield  {title} {\bibinfo {title} {Field-free superconducting diode effect in noncentrosymmetric superconductor/ferromagnet multilayers},\ }\href {https://www.nature.com/articles/s41565-022-01159-4} {\bibfield  {journal} {\bibinfo  {journal} {Nature Nanotechnology}\ }\textbf {\bibinfo {volume} {17}},\ \bibinfo {pages} {823} (\bibinfo {year}
  {2022})}\BibitemShut {NoStop}%
\bibitem [{\citenamefont {Jiang}\ \emph {et~al.}(2022)\citenamefont {Jiang}, \citenamefont {Milo{\v{s}}evi{\'c}}, \citenamefont {Wang}, \citenamefont {Xiao}, \citenamefont {Peeters},\ and\ \citenamefont {Chen}}]{jiang2022}%
  \BibitemOpen
  \bibfield  {author} {\bibinfo {author} {\bibfnamefont {J.}~\bibnamefont {Jiang}}, \bibinfo {author} {\bibfnamefont {M.}~\bibnamefont {Milo{\v{s}}evi{\'c}}}, \bibinfo {author} {\bibfnamefont {Y.-L.}\ \bibnamefont {Wang}}, \bibinfo {author} {\bibfnamefont {Z.-L.}\ \bibnamefont {Xiao}}, \bibinfo {author} {\bibfnamefont {F.}~\bibnamefont {Peeters}},\ and\ \bibinfo {author} {\bibfnamefont {Q.-H.}\ \bibnamefont {Chen}},\ }\bibfield  {title} {\bibinfo {title} {Field-free superconducting diode in a magnetically nanostructured superconductor},\ }\href {https://journals.aps.org/prapplied/abstract/10.1103/PhysRevApplied.18.034064} {\bibfield  {journal} {\bibinfo  {journal} {Physical Review Applied}\ }\textbf {\bibinfo {volume} {18}},\ \bibinfo {pages} {034064} (\bibinfo {year} {2022})}\BibitemShut {NoStop}%
\bibitem [{\citenamefont {Trahms}\ \emph {et~al.}(2023)\citenamefont {Trahms}, \citenamefont {Melischek}, \citenamefont {Steiner}, \citenamefont {Mahendru}, \citenamefont {Tamir}, \citenamefont {Bogdanoff}, \citenamefont {Peters}, \citenamefont {Reecht}, \citenamefont {Winkelmann}, \citenamefont {von Oppen} \emph {et~al.}}]{trahms2023}%
  \BibitemOpen
  \bibfield  {author} {\bibinfo {author} {\bibfnamefont {M.}~\bibnamefont {Trahms}}, \bibinfo {author} {\bibfnamefont {L.}~\bibnamefont {Melischek}}, \bibinfo {author} {\bibfnamefont {J.~F.}\ \bibnamefont {Steiner}}, \bibinfo {author} {\bibfnamefont {B.}~\bibnamefont {Mahendru}}, \bibinfo {author} {\bibfnamefont {I.}~\bibnamefont {Tamir}}, \bibinfo {author} {\bibfnamefont {N.}~\bibnamefont {Bogdanoff}}, \bibinfo {author} {\bibfnamefont {O.}~\bibnamefont {Peters}}, \bibinfo {author} {\bibfnamefont {G.}~\bibnamefont {Reecht}}, \bibinfo {author} {\bibfnamefont {C.~B.}\ \bibnamefont {Winkelmann}}, \bibinfo {author} {\bibfnamefont {F.}~\bibnamefont {von Oppen}}, \emph {et~al.},\ }\bibfield  {title} {\bibinfo {title} {Diode effect in josephson junctions with a single magnetic atom},\ }\href {https://www.nature.com/articles/s41586-023-05743-z} {\bibfield  {journal} {\bibinfo  {journal} {Nature}\ }\textbf {\bibinfo {volume} {615}},\ \bibinfo {pages} {628} (\bibinfo {year} {2023})}\BibitemShut {NoStop}%
\bibitem [{\citenamefont {Golod}\ and\ \citenamefont {Krasnov}(2022)}]{golod2022}%
  \BibitemOpen
  \bibfield  {author} {\bibinfo {author} {\bibfnamefont {T.}~\bibnamefont {Golod}}\ and\ \bibinfo {author} {\bibfnamefont {V.~M.}\ \bibnamefont {Krasnov}},\ }\bibfield  {title} {\bibinfo {title} {Demonstration of a superconducting diode-with-memory, operational at zero magnetic field with switchable nonreciprocity},\ }\href {https://www.nature.com/articles/s41467-022-31256-w} {\bibfield  {journal} {\bibinfo  {journal} {Nature Communications}\ }\textbf {\bibinfo {volume} {13}},\ \bibinfo {pages} {3658} (\bibinfo {year} {2022})}\BibitemShut {NoStop}%
\bibitem [{\citenamefont {Wakatsuki}\ \emph {et~al.}(2017)\citenamefont {Wakatsuki}, \citenamefont {Saito}, \citenamefont {Hoshino}, \citenamefont {Itahashi}, \citenamefont {Ideue}, \citenamefont {Ezawa}, \citenamefont {Iwasa},\ and\ \citenamefont {Nagaosa}}]{wakatsuki2017}%
  \BibitemOpen
  \bibfield  {author} {\bibinfo {author} {\bibfnamefont {R.}~\bibnamefont {Wakatsuki}}, \bibinfo {author} {\bibfnamefont {Y.}~\bibnamefont {Saito}}, \bibinfo {author} {\bibfnamefont {S.}~\bibnamefont {Hoshino}}, \bibinfo {author} {\bibfnamefont {Y.~M.}\ \bibnamefont {Itahashi}}, \bibinfo {author} {\bibfnamefont {T.}~\bibnamefont {Ideue}}, \bibinfo {author} {\bibfnamefont {M.}~\bibnamefont {Ezawa}}, \bibinfo {author} {\bibfnamefont {Y.}~\bibnamefont {Iwasa}},\ and\ \bibinfo {author} {\bibfnamefont {N.}~\bibnamefont {Nagaosa}},\ }\bibfield  {title} {\bibinfo {title} {Nonreciprocal charge transport in noncentrosymmetric superconductors},\ }\href {https://www.science.org/doi/10.1126/sciadv.1602390} {\bibfield  {journal} {\bibinfo  {journal} {Science Advances}\ }\textbf {\bibinfo {volume} {3}},\ \bibinfo {pages} {e1602390} (\bibinfo {year} {2017})}\BibitemShut {NoStop}%
\bibitem [{\citenamefont {Lin}\ \emph {et~al.}(2022)\citenamefont {Lin}, \citenamefont {Siriviboon}, \citenamefont {Scammell}, \citenamefont {Liu}, \citenamefont {Rhodes}, \citenamefont {Watanabe}, \citenamefont {Taniguchi}, \citenamefont {Hone}, \citenamefont {Scheurer},\ and\ \citenamefont {Li}}]{lin2022}%
  \BibitemOpen
  \bibfield  {author} {\bibinfo {author} {\bibfnamefont {J.-X.}\ \bibnamefont {Lin}}, \bibinfo {author} {\bibfnamefont {P.}~\bibnamefont {Siriviboon}}, \bibinfo {author} {\bibfnamefont {H.~D.}\ \bibnamefont {Scammell}}, \bibinfo {author} {\bibfnamefont {S.}~\bibnamefont {Liu}}, \bibinfo {author} {\bibfnamefont {D.}~\bibnamefont {Rhodes}}, \bibinfo {author} {\bibfnamefont {K.}~\bibnamefont {Watanabe}}, \bibinfo {author} {\bibfnamefont {T.}~\bibnamefont {Taniguchi}}, \bibinfo {author} {\bibfnamefont {J.}~\bibnamefont {Hone}}, \bibinfo {author} {\bibfnamefont {M.~S.}\ \bibnamefont {Scheurer}},\ and\ \bibinfo {author} {\bibfnamefont {J.}~\bibnamefont {Li}},\ }\bibfield  {title} {\bibinfo {title} {Zero-field superconducting diode effect in small-twist-angle trilayer graphene},\ }\href {https://www.nature.com/articles/s41567-022-01700-1} {\bibfield  {journal} {\bibinfo  {journal} {Nature Physics}\ }\textbf {\bibinfo {volume} {18}},\ \bibinfo {pages} {1221} (\bibinfo {year} {2022})}\BibitemShut {NoStop}%
\bibitem [{\citenamefont {Swartz}\ and\ \citenamefont {Hart~Jr}(1967)}]{swartz1967}%
  \BibitemOpen
  \bibfield  {author} {\bibinfo {author} {\bibfnamefont {P.}~\bibnamefont {Swartz}}\ and\ \bibinfo {author} {\bibfnamefont {H.}~\bibnamefont {Hart~Jr}},\ }\bibfield  {title} {\bibinfo {title} {Asymmetries of the critical surface current in type-ii superconductors},\ }\href {https://journals.aps.org/pr/abstract/10.1103/PhysRev.156.412} {\bibfield  {journal} {\bibinfo  {journal} {Physical Review}\ }\textbf {\bibinfo {volume} {156}},\ \bibinfo {pages} {412} (\bibinfo {year} {1967})}\BibitemShut {NoStop}%
\bibitem [{\citenamefont {Vodolazov}\ \emph {et~al.}(2005)\citenamefont {Vodolazov}, \citenamefont {Peeters}, \citenamefont {Grigorieva},\ and\ \citenamefont {Geim}}]{1vodolazov2005}%
  \BibitemOpen
  \bibfield  {author} {\bibinfo {author} {\bibfnamefont {D.}~\bibnamefont {Vodolazov}}, \bibinfo {author} {\bibfnamefont {F.}~\bibnamefont {Peeters}}, \bibinfo {author} {\bibfnamefont {I.}~\bibnamefont {Grigorieva}},\ and\ \bibinfo {author} {\bibfnamefont {A.}~\bibnamefont {Geim}},\ }\bibfield  {title} {\bibinfo {title} {Nonlocal response and surface-barrier-induced rectification in hall-shaped mesoscopic superconductors},\ }\href {https://journals.aps.org/prb/abstract/10.1103/PhysRevB.72.024537} {\bibfield  {journal} {\bibinfo  {journal} {Physical Review B}\ }\textbf {\bibinfo {volume} {72}},\ \bibinfo {pages} {024537} (\bibinfo {year} {2005})}\BibitemShut {NoStop}%
\bibitem [{\citenamefont {Vodolazov}\ and\ \citenamefont {Peeters}(2005)}]{2vodolazov2005}%
  \BibitemOpen
  \bibfield  {author} {\bibinfo {author} {\bibfnamefont {D.~Y.}\ \bibnamefont {Vodolazov}}\ and\ \bibinfo {author} {\bibfnamefont {F.}~\bibnamefont {Peeters}},\ }\bibfield  {title} {\bibinfo {title} {Superconducting rectifier based on the asymmetric surface barrier effect},\ }\href {https://journals.aps.org/prb/abstract/10.1103/PhysRevB.72.172508} {\bibfield  {journal} {\bibinfo  {journal} {Physical Review B}\ }\textbf {\bibinfo {volume} {72}},\ \bibinfo {pages} {172508} (\bibinfo {year} {2005})}\BibitemShut {NoStop}%
\bibitem [{\citenamefont {Moll}\ and\ \citenamefont {Geshkenbein}(2023)}]{moll2023}%
  \BibitemOpen
  \bibfield  {author} {\bibinfo {author} {\bibfnamefont {P.~J.}\ \bibnamefont {Moll}}\ and\ \bibinfo {author} {\bibfnamefont {V.~B.}\ \bibnamefont {Geshkenbein}},\ }\bibfield  {title} {\bibinfo {title} {Evolution of superconducting diodes},\ }\href {https://www.nature.com/articles/s41567-023-02229-7} {\bibfield  {journal} {\bibinfo  {journal} {Nature Physics}\ }\textbf {\bibinfo {volume} {19}},\ \bibinfo {pages} {1379} (\bibinfo {year} {2023})}\BibitemShut {NoStop}%
\bibitem [{\citenamefont {Edwards}\ and\ \citenamefont {Newhouse}(1962)}]{edwards1962}%
  \BibitemOpen
  \bibfield  {author} {\bibinfo {author} {\bibfnamefont {H.}~\bibnamefont {Edwards}}\ and\ \bibinfo {author} {\bibfnamefont {V.}~\bibnamefont {Newhouse}},\ }\bibfield  {title} {\bibinfo {title} {Superconducting film geometry with strong critical current asymmetry},\ }\href {https://pubs.aip.org/aip/jap/article-abstract/33/3/868/1804/Superconducting-Film-Geometry-With-Strong-Critical?redirectedFrom=fulltext} {\bibfield  {journal} {\bibinfo  {journal} {Journal of Applied Physics}\ }\textbf {\bibinfo {volume} {33}},\ \bibinfo {pages} {868} (\bibinfo {year} {1962})}\BibitemShut {NoStop}%
\bibitem [{\citenamefont {Yasuda}\ \emph {et~al.}(2019)\citenamefont {Yasuda}, \citenamefont {Yasuda}, \citenamefont {Liang}, \citenamefont {Yoshimi}, \citenamefont {Tsukazaki}, \citenamefont {Takahashi}, \citenamefont {Nagaosa}, \citenamefont {Kawasaki},\ and\ \citenamefont {Tokura}}]{yasuda2019}%
  \BibitemOpen
  \bibfield  {author} {\bibinfo {author} {\bibfnamefont {K.}~\bibnamefont {Yasuda}}, \bibinfo {author} {\bibfnamefont {H.}~\bibnamefont {Yasuda}}, \bibinfo {author} {\bibfnamefont {T.}~\bibnamefont {Liang}}, \bibinfo {author} {\bibfnamefont {R.}~\bibnamefont {Yoshimi}}, \bibinfo {author} {\bibfnamefont {A.}~\bibnamefont {Tsukazaki}}, \bibinfo {author} {\bibfnamefont {K.~S.}\ \bibnamefont {Takahashi}}, \bibinfo {author} {\bibfnamefont {N.}~\bibnamefont {Nagaosa}}, \bibinfo {author} {\bibfnamefont {M.}~\bibnamefont {Kawasaki}},\ and\ \bibinfo {author} {\bibfnamefont {Y.}~\bibnamefont {Tokura}},\ }\bibfield  {title} {\bibinfo {title} {Nonreciprocal charge transport at topological insulator/superconductor interface},\ }\href {https://www.nature.com/articles/s41535-022-00514-x} {\bibfield  {journal} {\bibinfo  {journal} {Nature Communications}\ }\textbf {\bibinfo {volume} {10}},\ \bibinfo {pages} {2734} (\bibinfo {year} {2019})}\BibitemShut {NoStop}%
\bibitem [{\citenamefont {Itahashi}\ \emph {et~al.}(2020)\citenamefont {Itahashi}, \citenamefont {Ideue}, \citenamefont {Saito}, \citenamefont {Shimizu}, \citenamefont {Ouchi}, \citenamefont {Nojima},\ and\ \citenamefont {Iwasa}}]{itahashi2020}%
  \BibitemOpen
  \bibfield  {author} {\bibinfo {author} {\bibfnamefont {Y.~M.}\ \bibnamefont {Itahashi}}, \bibinfo {author} {\bibfnamefont {T.}~\bibnamefont {Ideue}}, \bibinfo {author} {\bibfnamefont {Y.}~\bibnamefont {Saito}}, \bibinfo {author} {\bibfnamefont {S.}~\bibnamefont {Shimizu}}, \bibinfo {author} {\bibfnamefont {T.}~\bibnamefont {Ouchi}}, \bibinfo {author} {\bibfnamefont {T.}~\bibnamefont {Nojima}},\ and\ \bibinfo {author} {\bibfnamefont {Y.}~\bibnamefont {Iwasa}},\ }\bibfield  {title} {\bibinfo {title} {{Nonreciprocal transport in gate-induced polar superconductor SrTiO3}},\ }\href {https://www.science.org/doi/full/10.1126/sciadv.aay9120} {\bibfield  {journal} {\bibinfo  {journal} {Science advances}\ }\textbf {\bibinfo {volume} {6}},\ \bibinfo {pages} {eaay9120} (\bibinfo {year} {2020})}\BibitemShut {NoStop}%
\bibitem [{\citenamefont {Wu}\ \emph {et~al.}(2022)\citenamefont {Wu}, \citenamefont {Wang}, \citenamefont {Xu}, \citenamefont {Sivakumar}, \citenamefont {Pasco}, \citenamefont {Filippozzi}, \citenamefont {Parkin}, \citenamefont {Zeng}, \citenamefont {McQueen},\ and\ \citenamefont {Ali}}]{wu2022}%
  \BibitemOpen
  \bibfield  {author} {\bibinfo {author} {\bibfnamefont {H.}~\bibnamefont {Wu}}, \bibinfo {author} {\bibfnamefont {Y.}~\bibnamefont {Wang}}, \bibinfo {author} {\bibfnamefont {Y.}~\bibnamefont {Xu}}, \bibinfo {author} {\bibfnamefont {P.~K.}\ \bibnamefont {Sivakumar}}, \bibinfo {author} {\bibfnamefont {C.}~\bibnamefont {Pasco}}, \bibinfo {author} {\bibfnamefont {U.}~\bibnamefont {Filippozzi}}, \bibinfo {author} {\bibfnamefont {S.~S.}\ \bibnamefont {Parkin}}, \bibinfo {author} {\bibfnamefont {Y.-J.}\ \bibnamefont {Zeng}}, \bibinfo {author} {\bibfnamefont {T.}~\bibnamefont {McQueen}},\ and\ \bibinfo {author} {\bibfnamefont {M.~N.}\ \bibnamefont {Ali}},\ }\bibfield  {title} {\bibinfo {title} {The field-free josephson diode in a van der waals heterostructure},\ }\href {https://www.nature.com/articles/s41586-022-04504-8} {\bibfield  {journal} {\bibinfo  {journal} {Nature}\ }\textbf {\bibinfo {volume} {604}},\ \bibinfo {pages} {653} (\bibinfo {year} {2022})}\BibitemShut {NoStop}%
\bibitem [{\citenamefont {Jeon}\ \emph {et~al.}(2022)\citenamefont {Jeon}, \citenamefont {Kim}, \citenamefont {Yoon}, \citenamefont {Jeon}, \citenamefont {Han}, \citenamefont {Cottet}, \citenamefont {Kontos},\ and\ \citenamefont {Parkin}}]{jeon2022}%
  \BibitemOpen
  \bibfield  {author} {\bibinfo {author} {\bibfnamefont {K.-R.}\ \bibnamefont {Jeon}}, \bibinfo {author} {\bibfnamefont {J.-K.}\ \bibnamefont {Kim}}, \bibinfo {author} {\bibfnamefont {J.}~\bibnamefont {Yoon}}, \bibinfo {author} {\bibfnamefont {J.-C.}\ \bibnamefont {Jeon}}, \bibinfo {author} {\bibfnamefont {H.}~\bibnamefont {Han}}, \bibinfo {author} {\bibfnamefont {A.}~\bibnamefont {Cottet}}, \bibinfo {author} {\bibfnamefont {T.}~\bibnamefont {Kontos}},\ and\ \bibinfo {author} {\bibfnamefont {S.~S.}\ \bibnamefont {Parkin}},\ }\bibfield  {title} {\bibinfo {title} {Zero-field polarity-reversible josephson supercurrent diodes enabled by a proximity-magnetized pt barrier},\ }\href {https://www.nature.com/articles/s41563-022-01300-7} {\bibfield  {journal} {\bibinfo  {journal} {Nature Materials}\ }\textbf {\bibinfo {volume} {21}},\ \bibinfo {pages} {1008} (\bibinfo {year} {2022})}\BibitemShut {NoStop}%
\bibitem [{\citenamefont {Baumgartner}\ \emph {et~al.}(2022)\citenamefont {Baumgartner}, \citenamefont {Fuchs}, \citenamefont {Costa}, \citenamefont {Reinhardt}, \citenamefont {Gronin}, \citenamefont {Gardner}, \citenamefont {Lindemann}, \citenamefont {Manfra}, \citenamefont {Faria~Junior}, \citenamefont {Kochan} \emph {et~al.}}]{baumgartner2022}%
  \BibitemOpen
  \bibfield  {author} {\bibinfo {author} {\bibfnamefont {C.}~\bibnamefont {Baumgartner}}, \bibinfo {author} {\bibfnamefont {L.}~\bibnamefont {Fuchs}}, \bibinfo {author} {\bibfnamefont {A.}~\bibnamefont {Costa}}, \bibinfo {author} {\bibfnamefont {S.}~\bibnamefont {Reinhardt}}, \bibinfo {author} {\bibfnamefont {S.}~\bibnamefont {Gronin}}, \bibinfo {author} {\bibfnamefont {G.~C.}\ \bibnamefont {Gardner}}, \bibinfo {author} {\bibfnamefont {T.}~\bibnamefont {Lindemann}}, \bibinfo {author} {\bibfnamefont {M.~J.}\ \bibnamefont {Manfra}}, \bibinfo {author} {\bibfnamefont {P.~E.}\ \bibnamefont {Faria~Junior}}, \bibinfo {author} {\bibfnamefont {D.}~\bibnamefont {Kochan}}, \emph {et~al.},\ }\bibfield  {title} {\bibinfo {title} {Supercurrent rectification and magnetochiral effects in symmetric josephson junctions},\ }\href {https://www.nature.com/articles/s41565-021-01009-9} {\bibfield  {journal} {\bibinfo  {journal} {Nature Nanotechnology}\ }\textbf {\bibinfo {volume} {17}},\ \bibinfo {pages} {39} (\bibinfo {year}
  {2022})}\BibitemShut {NoStop}%
\bibitem [{\citenamefont {Ili{\'c}}\ and\ \citenamefont {Bergeret}(2022)}]{ilic2022}%
  \BibitemOpen
  \bibfield  {author} {\bibinfo {author} {\bibfnamefont {S.}~\bibnamefont {Ili{\'c}}}\ and\ \bibinfo {author} {\bibfnamefont {F.~S.}\ \bibnamefont {Bergeret}},\ }\bibfield  {title} {\bibinfo {title} {Theory of the supercurrent diode effect in rashba superconductors with arbitrary disorder},\ }\href {https://journals.aps.org/prl/abstract/10.1103/PhysRevLett.128.177001} {\bibfield  {journal} {\bibinfo  {journal} {Physical Review Letters}\ }\textbf {\bibinfo {volume} {128}},\ \bibinfo {pages} {177001} (\bibinfo {year} {2022})}\BibitemShut {NoStop}%
\bibitem [{\citenamefont {Kokkeler}\ \emph {et~al.}(2022)\citenamefont {Kokkeler}, \citenamefont {Golubov},\ and\ \citenamefont {Bergeret}}]{kokkeler2022}%
  \BibitemOpen
  \bibfield  {author} {\bibinfo {author} {\bibfnamefont {T.}~\bibnamefont {Kokkeler}}, \bibinfo {author} {\bibfnamefont {A.}~\bibnamefont {Golubov}},\ and\ \bibinfo {author} {\bibfnamefont {F.}~\bibnamefont {Bergeret}},\ }\bibfield  {title} {\bibinfo {title} {Field-free anomalous junction and superconducting diode effect in spin-split superconductor/topological insulator junctions},\ }\href {https://journals.aps.org/prb/abstract/10.1103/PhysRevB.106.214504} {\bibfield  {journal} {\bibinfo  {journal} {Physical Review B}\ }\textbf {\bibinfo {volume} {106}},\ \bibinfo {pages} {214504} (\bibinfo {year} {2022})}\BibitemShut {NoStop}%
\bibitem [{\citenamefont {Yuan}\ and\ \citenamefont {Fu}(2022)}]{yuan2022}%
  \BibitemOpen
  \bibfield  {author} {\bibinfo {author} {\bibfnamefont {N.~F.}\ \bibnamefont {Yuan}}\ and\ \bibinfo {author} {\bibfnamefont {L.}~\bibnamefont {Fu}},\ }\bibfield  {title} {\bibinfo {title} {Supercurrent diode effect and finite-momentum superconductors},\ }\href {https://www.pnas.org/doi/full/10.1073/pnas.2119548119} {\bibfield  {journal} {\bibinfo  {journal} {Proceedings of the National Academy of Sciences}\ }\textbf {\bibinfo {volume} {119}},\ \bibinfo {pages} {e2119548119} (\bibinfo {year} {2022})}\BibitemShut {NoStop}%
\bibitem [{\citenamefont {Davydova}\ \emph {et~al.}(2022)\citenamefont {Davydova}, \citenamefont {Prembabu},\ and\ \citenamefont {Fu}}]{davydova2022}%
  \BibitemOpen
  \bibfield  {author} {\bibinfo {author} {\bibfnamefont {M.}~\bibnamefont {Davydova}}, \bibinfo {author} {\bibfnamefont {S.}~\bibnamefont {Prembabu}},\ and\ \bibinfo {author} {\bibfnamefont {L.}~\bibnamefont {Fu}},\ }\bibfield  {title} {\bibinfo {title} {Universal josephson diode effect},\ }\href {https://www.science.org/doi/10.1126/sciadv.abo0309} {\bibfield  {journal} {\bibinfo  {journal} {Science advances}\ }\textbf {\bibinfo {volume} {8}},\ \bibinfo {pages} {eabo0309} (\bibinfo {year} {2022})}\BibitemShut {NoStop}%
\bibitem [{\citenamefont {Pal}\ \emph {et~al.}(2022)\citenamefont {Pal}, \citenamefont {Chakraborty}, \citenamefont {Sivakumar}, \citenamefont {Davydova}, \citenamefont {Gopi}, \citenamefont {Pandeya}, \citenamefont {Krieger}, \citenamefont {Zhang}, \citenamefont {Date}, \citenamefont {Ju} \emph {et~al.}}]{pal2022}%
  \BibitemOpen
  \bibfield  {author} {\bibinfo {author} {\bibfnamefont {B.}~\bibnamefont {Pal}}, \bibinfo {author} {\bibfnamefont {A.}~\bibnamefont {Chakraborty}}, \bibinfo {author} {\bibfnamefont {P.~K.}\ \bibnamefont {Sivakumar}}, \bibinfo {author} {\bibfnamefont {M.}~\bibnamefont {Davydova}}, \bibinfo {author} {\bibfnamefont {A.~K.}\ \bibnamefont {Gopi}}, \bibinfo {author} {\bibfnamefont {A.~K.}\ \bibnamefont {Pandeya}}, \bibinfo {author} {\bibfnamefont {J.~A.}\ \bibnamefont {Krieger}}, \bibinfo {author} {\bibfnamefont {Y.}~\bibnamefont {Zhang}}, \bibinfo {author} {\bibfnamefont {M.}~\bibnamefont {Date}}, \bibinfo {author} {\bibfnamefont {S.}~\bibnamefont {Ju}}, \emph {et~al.},\ }\bibfield  {title} {\bibinfo {title} {Josephson diode effect from cooper pair momentum in a topological semimetal},\ }\href {https://www.nature.com/articles/s41567-022-01699-5} {\bibfield  {journal} {\bibinfo  {journal} {Nature Physics}\ }\textbf {\bibinfo {volume} {18}},\ \bibinfo {pages} {1228} (\bibinfo {year} {2022})}\BibitemShut {NoStop}%
\bibitem [{\citenamefont {Hope}\ \emph {et~al.}(2021)\citenamefont {Hope}, \citenamefont {Amundsen}, \citenamefont {Suri}, \citenamefont {Moodera},\ and\ \citenamefont {Kamra}}]{hope2021}%
  \BibitemOpen
  \bibfield  {author} {\bibinfo {author} {\bibfnamefont {M.~K.}\ \bibnamefont {Hope}}, \bibinfo {author} {\bibfnamefont {M.}~\bibnamefont {Amundsen}}, \bibinfo {author} {\bibfnamefont {D.}~\bibnamefont {Suri}}, \bibinfo {author} {\bibfnamefont {J.~S.}\ \bibnamefont {Moodera}},\ and\ \bibinfo {author} {\bibfnamefont {A.}~\bibnamefont {Kamra}},\ }\bibfield  {title} {\bibinfo {title} {Interfacial control of vortex-limited critical current in type-ii superconductor films},\ }\href {https://journals.aps.org/prb/abstract/10.1103/PhysRevB.104.184512} {\bibfield  {journal} {\bibinfo  {journal} {Physical Review B}\ }\textbf {\bibinfo {volume} {104}},\ \bibinfo {pages} {184512} (\bibinfo {year} {2021})}\BibitemShut {NoStop}%
\bibitem [{\citenamefont {Zhao}\ \emph {et~al.}(2023)\citenamefont {Zhao}, \citenamefont {Cui}, \citenamefont {Volkov}, \citenamefont {Yoo}, \citenamefont {Lee}, \citenamefont {Gardener}, \citenamefont {Akey}, \citenamefont {Engelke}, \citenamefont {Ronen}, \citenamefont {Zhong} \emph {et~al.}}]{zhao2023}%
  \BibitemOpen
  \bibfield  {author} {\bibinfo {author} {\bibfnamefont {S.~F.}\ \bibnamefont {Zhao}}, \bibinfo {author} {\bibfnamefont {X.}~\bibnamefont {Cui}}, \bibinfo {author} {\bibfnamefont {P.~A.}\ \bibnamefont {Volkov}}, \bibinfo {author} {\bibfnamefont {H.}~\bibnamefont {Yoo}}, \bibinfo {author} {\bibfnamefont {S.}~\bibnamefont {Lee}}, \bibinfo {author} {\bibfnamefont {J.~A.}\ \bibnamefont {Gardener}}, \bibinfo {author} {\bibfnamefont {A.~J.}\ \bibnamefont {Akey}}, \bibinfo {author} {\bibfnamefont {R.}~\bibnamefont {Engelke}}, \bibinfo {author} {\bibfnamefont {Y.}~\bibnamefont {Ronen}}, \bibinfo {author} {\bibfnamefont {R.}~\bibnamefont {Zhong}}, \emph {et~al.},\ }\bibfield  {title} {\bibinfo {title} {Time-reversal symmetry breaking superconductivity between twisted cuprate superconductors},\ }\href {https://www.science.org/doi/abs/10.1126/science.abl8371} {\bibfield  {journal} {\bibinfo  {journal} {Science}\ }\textbf {\bibinfo {volume} {382}},\ \bibinfo {pages} {1422} (\bibinfo {year} {2023})}\BibitemShut {NoStop}%
\bibitem [{\citenamefont {Ghosh}\ \emph {et~al.}(2024)\citenamefont {Ghosh}, \citenamefont {Patil}, \citenamefont {Basu}, \citenamefont {Kuldeep}, \citenamefont {Dutta}, \citenamefont {Jangade}, \citenamefont {Kulkarni}, \citenamefont {Thamizhavel}, \citenamefont {Steiner}, \citenamefont {von Oppen} \emph {et~al.}}]{ghosh2024}%
  \BibitemOpen
  \bibfield  {author} {\bibinfo {author} {\bibfnamefont {S.}~\bibnamefont {Ghosh}}, \bibinfo {author} {\bibfnamefont {V.}~\bibnamefont {Patil}}, \bibinfo {author} {\bibfnamefont {A.}~\bibnamefont {Basu}}, \bibinfo {author} {\bibnamefont {Kuldeep}}, \bibinfo {author} {\bibfnamefont {A.}~\bibnamefont {Dutta}}, \bibinfo {author} {\bibfnamefont {D.~A.}\ \bibnamefont {Jangade}}, \bibinfo {author} {\bibfnamefont {R.}~\bibnamefont {Kulkarni}}, \bibinfo {author} {\bibfnamefont {A.}~\bibnamefont {Thamizhavel}}, \bibinfo {author} {\bibfnamefont {J.~F.}\ \bibnamefont {Steiner}}, \bibinfo {author} {\bibfnamefont {F.}~\bibnamefont {von Oppen}}, \emph {et~al.},\ }\bibfield  {title} {\bibinfo {title} {High-temperature josephson diode},\ }\href {https://www.nature.com/articles/s41563-024-01804-4} {\bibfield  {journal} {\bibinfo  {journal} {Nature Materials}\ }\textbf {\bibinfo {volume} {23}},\ \bibinfo {pages} {612–618} (\bibinfo {year} {2024})}\BibitemShut {NoStop}%
\bibitem [{\citenamefont {Bozkurt}\ \emph {et~al.}(2023)\citenamefont {Bozkurt}, \citenamefont {Brookman}, \citenamefont {Fatemi},\ and\ \citenamefont {Akhmerov}}]{bozkurt2023}%
  \BibitemOpen
  \bibfield  {author} {\bibinfo {author} {\bibfnamefont {A.~M.}\ \bibnamefont {Bozkurt}}, \bibinfo {author} {\bibfnamefont {J.}~\bibnamefont {Brookman}}, \bibinfo {author} {\bibfnamefont {V.}~\bibnamefont {Fatemi}},\ and\ \bibinfo {author} {\bibfnamefont {A.~R.}\ \bibnamefont {Akhmerov}},\ }\bibfield  {title} {\bibinfo {title} {Double-fourier engineering of josephson energy-phase relationships applied to diodes},\ }\href {https://scipost.org/SciPostPhys.15.5.204} {\bibfield  {journal} {\bibinfo  {journal} {SciPost Physics}\ }\textbf {\bibinfo {volume} {15}},\ \bibinfo {pages} {204} (\bibinfo {year} {2023})}\BibitemShut {NoStop}%
\bibitem [{\citenamefont {Cayao}\ \emph {et~al.}(2024)\citenamefont {Cayao}, \citenamefont {Nagaosa},\ and\ \citenamefont {Tanaka}}]{cayao2024}%
  \BibitemOpen
  \bibfield  {author} {\bibinfo {author} {\bibfnamefont {J.}~\bibnamefont {Cayao}}, \bibinfo {author} {\bibfnamefont {N.}~\bibnamefont {Nagaosa}},\ and\ \bibinfo {author} {\bibfnamefont {Y.}~\bibnamefont {Tanaka}},\ }\bibfield  {title} {\bibinfo {title} {Enhancing the josephson diode effect with majorana bound states},\ }\href {https://journals.aps.org/prb/abstract/10.1103/PhysRevB.109.L081405} {\bibfield  {journal} {\bibinfo  {journal} {Physical Review B}\ }\textbf {\bibinfo {volume} {109}},\ \bibinfo {pages} {L081405} (\bibinfo {year} {2024})}\BibitemShut {NoStop}%
\bibitem [{\citenamefont {Yerin}\ \emph {et~al.}(2024)\citenamefont {Yerin}, \citenamefont {Drechsler}, \citenamefont {Varlamov}, \citenamefont {Cuoco},\ and\ \citenamefont {Giazotto}}]{yerin2024}%
  \BibitemOpen
  \bibfield  {author} {\bibinfo {author} {\bibfnamefont {Y.}~\bibnamefont {Yerin}}, \bibinfo {author} {\bibfnamefont {S.-L.}\ \bibnamefont {Drechsler}}, \bibinfo {author} {\bibfnamefont {A.}~\bibnamefont {Varlamov}}, \bibinfo {author} {\bibfnamefont {M.}~\bibnamefont {Cuoco}},\ and\ \bibinfo {author} {\bibfnamefont {F.}~\bibnamefont {Giazotto}},\ }\bibfield  {title} {\bibinfo {title} {Supercurrent rectification with time-reversal symmetry broken multiband superconductors},\ }\href {https://arxiv.org/abs/2404.12641} {\bibfield  {journal} {\bibinfo  {journal} {arXiv preprint arXiv:2404.12641}\ } (\bibinfo {year} {2024})}\BibitemShut {NoStop}%
\bibitem [{\citenamefont {Pal}\ and\ \citenamefont {Benjamin}(2019)}]{pal2018}%
  \BibitemOpen
  \bibfield  {author} {\bibinfo {author} {\bibfnamefont {S.}~\bibnamefont {Pal}}\ and\ \bibinfo {author} {\bibfnamefont {C.}~\bibnamefont {Benjamin}},\ }\bibfield  {title} {\bibinfo {title} {Quantized josephson phase battery},\ }\href {https://iopscience.iop.org/article/10.1209/0295-5075/126/57002} {\bibfield  {journal} {\bibinfo  {journal} {Europhysics Letters}\ }\textbf {\bibinfo {volume} {126}},\ \bibinfo {pages} {57002} (\bibinfo {year} {2019})}\BibitemShut {NoStop}%
\bibitem [{\citenamefont {Horowitz}\ \emph {et~al.}(1989)\citenamefont {Horowitz}, \citenamefont {Hill},\ and\ \citenamefont {Robinson}}]{horowitz1989}%
  \BibitemOpen
  \bibfield  {author} {\bibinfo {author} {\bibfnamefont {P.}~\bibnamefont {Horowitz}}, \bibinfo {author} {\bibfnamefont {W.}~\bibnamefont {Hill}},\ and\ \bibinfo {author} {\bibfnamefont {I.}~\bibnamefont {Robinson}},\ }\href@noop {} {\emph {\bibinfo {title} {The art of electronics}}},\ Vol.~\bibinfo {volume} {2}\ (\bibinfo  {publisher} {Cambridge university press Cambridge},\ \bibinfo {year} {1989})\BibitemShut {NoStop}%
\bibitem [{\citenamefont {Chapman}\ \emph {et~al.}(2017)\citenamefont {Chapman}, \citenamefont {Rosenthal}, \citenamefont {Kerckhoff}, \citenamefont {Moores}, \citenamefont {Vale}, \citenamefont {Mates}, \citenamefont {Hilton}, \citenamefont {Lalumiere}, \citenamefont {Blais},\ and\ \citenamefont {Lehnert}}]{chapman2017}%
  \BibitemOpen
  \bibfield  {author} {\bibinfo {author} {\bibfnamefont {B.~J.}\ \bibnamefont {Chapman}}, \bibinfo {author} {\bibfnamefont {E.~I.}\ \bibnamefont {Rosenthal}}, \bibinfo {author} {\bibfnamefont {J.}~\bibnamefont {Kerckhoff}}, \bibinfo {author} {\bibfnamefont {B.~A.}\ \bibnamefont {Moores}}, \bibinfo {author} {\bibfnamefont {L.~R.}\ \bibnamefont {Vale}}, \bibinfo {author} {\bibfnamefont {J.}~\bibnamefont {Mates}}, \bibinfo {author} {\bibfnamefont {G.~C.}\ \bibnamefont {Hilton}}, \bibinfo {author} {\bibfnamefont {K.}~\bibnamefont {Lalumiere}}, \bibinfo {author} {\bibfnamefont {A.}~\bibnamefont {Blais}},\ and\ \bibinfo {author} {\bibfnamefont {K.}~\bibnamefont {Lehnert}},\ }\bibfield  {title} {\bibinfo {title} {Widely tunable on-chip microwave circulator for superconducting quantum circuits},\ }\href {https://journals.aps.org/prx/abstract/10.1103/PhysRevX.7.041043} {\bibfield  {journal} {\bibinfo  {journal} {Physical Review X}\ }\textbf {\bibinfo {volume} {7}},\ \bibinfo {pages} {041043} (\bibinfo {year}
  {2017})}\BibitemShut {NoStop}%
\bibitem [{\citenamefont {Castellani}\ \emph {et~al.}(2024)\citenamefont {Castellani}, \citenamefont {Medeiros}, \citenamefont {Buzzi}, \citenamefont {Foster}, \citenamefont {Colangelo},\ and\ \citenamefont {Berggren}}]{castellani2024}%
  \BibitemOpen
  \bibfield  {author} {\bibinfo {author} {\bibfnamefont {M.}~\bibnamefont {Castellani}}, \bibinfo {author} {\bibfnamefont {O.}~\bibnamefont {Medeiros}}, \bibinfo {author} {\bibfnamefont {A.}~\bibnamefont {Buzzi}}, \bibinfo {author} {\bibfnamefont {R.~A.}\ \bibnamefont {Foster}}, \bibinfo {author} {\bibfnamefont {M.}~\bibnamefont {Colangelo}},\ and\ \bibinfo {author} {\bibfnamefont {K.~K.}\ \bibnamefont {Berggren}},\ }\bibfield  {title} {\bibinfo {title} {A superconducting full-wave bridge rectifier},\ }\href {https://arxiv.org/abs/2406.12175} {\bibfield  {journal} {\bibinfo  {journal} {arXiv preprint arXiv:2406.12175}\ } (\bibinfo {year} {2024})}\BibitemShut {NoStop}%
\end{thebibliography}%

\end{document}